\documentclass{segabs}
\usepackage{lipsum}
\hypersetup{
}

\begin{document}

\title{First arrival picking using U-net with Lovasz loss and nearest point picking method}

\renewcommand{\thefootnote}{\fnsymbol{footnote}} 

\author{Pengyu Yuan, University of Houston; Wenyi Hu, Advanced Geophysical Technology, Inc; 
Xuqing Wu, Jiefu Chen, Hien Van Nguyen, University of Houston}

\righthead{First arrival picking}

\maketitle

\begin{abstract}
We proposed a robust segmentation and picking workflow to solve the first arrival picking problem for seismic signal processing. Unlike traditional classification algorithm, image segmentation method can utilize the location information by outputting a prediction map which has the same size of the input image. A parameter-free nearest point picking algorithm is proposed to further improve the accuracy of the first arrival picking. The algorithm is test on synthetic clean data, synthetic noisy data, synthetic picking-disconnected data and field data. It performs well on all of them and the picking deviation reaches as low as 4.8ms per receiver.

The first arrival picking problem is formulated as the contour detection problem. Similar to \cite{wu2019semi}, we use U-net to perform the segmentation as it is proven to be state-of-the-art in many image segmentation tasks. Particularly, a Lovasz loss instead of the traditional cross-entropy loss is used to train the network for a better segmentation performance. Lovasz loss is a surrogate loss for Jaccard index or the so-called intersection-over-union (IoU) score, which is often one of the most used metrics for segmentation tasks. In the picking part, we use a novel nearest point picking (NPP) method  to take the advantage of the coherence of the first arrival picking among adjacent receivers. Our model is tested and validated on both synthetic and field data with harmonic noises.
The main contributions of this paper are as follows:
\begin{itemize}
    \item Used Lovasz loss to directly optimize the IoU for segmentation task. Improvement over the cross-entropy loss with regard to the segmentation accuracy is verified by the test result.
    \item Proposed a nearest point picking post processing method to overcome any defects left by the segmentation output.
    \item Conducted noise analysis and verified the model with both noisy synthetic and field datasets.
\end{itemize}
\end{abstract}

\section{Introduction}

In the seismic data processing area, first arrival picking is detecting the first arrivals of refracted signals from all signals received by receiver arrays. It is important but challenging because it is strongly affected by the subsurface structure, signal-to-noise ratio, etc. First arrival picking requires extensive manual interventions by domain experts, which is labor-intensive and time consuming. 

Manual picking is a straightforward method but it depends on the experience of experts and the estimation criteria is highly subjective. If the data volume is large, picking may consume a lot of time. The semi-automatic picking method is developed in order to pick the first arrivals faster(\cite{peraldi1972digital}). However, they tend to fail when the pulse shape changes from trace to trace and when bad or dead traces appear . Automatic first arrival picking methods are always preferred and a variety of these kind of techniques are explored over the decades. Energy ratio based methods may perform badly when the data quality is not good enough(\cite{coppens1985first} \cite{chen2005multi} \cite{wong2009automatic}). Another branch of first arrival picking method is to use neural networks. \cite{mccormack1993first}, \cite{maity2014novel}, and \cite{mousavi2016seismic} used the shallow neural network and these methods are highly sensitive to hand-designed features such as STA/LTA ratio, the amplitude, and the frequency. 

In recent years, deep learning has achieved great success in many fields, especially for computer vision. The seismic data are also images. \cite{hollander2018using}, \cite{yuan2018seismic}, and \cite{duan2018integrating} use deep neural networks as a classifier, to decide whether there is a first arrival signal or it is a poor picking. Both \cite{tsai2018first} and \cite{wu2019semi} formulated the first arrival picking problem as the segmentation problem. It takes advantage of the global information of the image by splitting the image into two parts. The first arrival we want is the edge between the signal area and non-signal area. However, segmentation alone cannot guarantee that the picking results meet the minimum accuracy requirement and post-processing is necessary to fix the residual segmentation defects. A nearest point picking post-processing method is proposed in this paper and it significantly improves the overall performance of the workflow.

\section{Problem Description}
\renewcommand{\figdir}{Fig} 
\plot{prob}{width=\columnwidth}
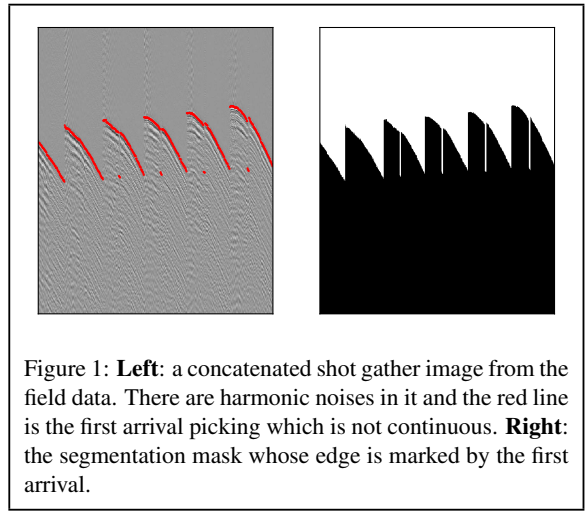
{\textbf{Left}: a concatenated shot gather image from the field data. There are harmonic noises in it and the red line is the first arrival picking which is not continuous. \textbf{Right}: the segmentation mask whose edge is marked by the first arrival.}

\autoref{fig:prob} shows a concatenated shot gather image from the field data and the ground truth segmentation mask. The horizontal axis is different receivers while the vertical axis represents the time. Each column in the image is a trace, representing the response of the seismic signal from a source to a receiver throughout the time steps. The task is to segment the seismic image into two parts and then pick the boundary automatically. 

In the ideal case, before the first arrival signal, the receiver should receive no signals. However this is not the case in the field data. The original seismic trace is contaminated by multiple types of noises and they exists in the whole trace. The picking line may also be disconnected because of missing traces or because we want to pick first arrivals for several shots at the same time. The number of receivers is not consistent for each shot gathers. Because harmonic noise radiated from power-line is very typical in seismic data, In this paper, we simulated the harmonic noises and disconnected situations on synthetic data and also tested on the field data to verify the robustness of the proposed model and workflow.

\section{Methodology}
\subsection{U-net}
\cite{ronneberger2015u} proposed a convolutional network called U-net with contracting path and expansive path to output segmentation map which has similar dimension with the input image. Both high-level and low-level features are extracted by the downsampling path, the upsampling path, and the skip connections. As shown in \autoref{fig:Unet}, the contracting path is composed of 3 blocks. First two blocks are composed of two convolution layers followed by one max pooling layer. The last block only has two convolution layers and it also acts as a bridge to the decoder part. Each convolution layer uses $3 \times 3$ kernels. Batch normalization and Relu activation function are used after the convolution layer. The decoder is symmetric to the encoder and it has 2 blocks. Each block is constructed by two convolution layers and one unpooling layer. At the end of the decoder, a $1 \times 1$ convolution layer is added to project the 32 feature maps to only 2 classes. And a softmax function is used to map the unnormalized logits to the probabilities. The output segmentation map assigns each single pixel to an area which has or does not have seismic signals. 
\plot{Unet}{width=\columnwidth}
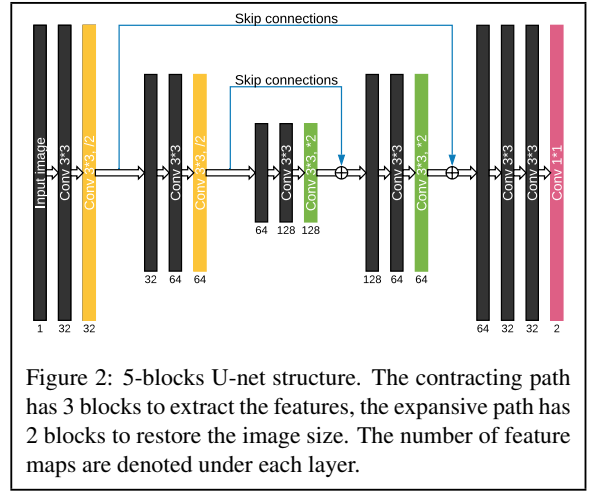
{5-blocks U-net structure. The contracting path has 3 blocks to extract the features, the expansive path has 2 blocks to restore the image size. The number of feature maps are denoted under each layer.}

\subsection{Loss function}
\subsubsection{Cross-entropy}
One of the most widely used functions in the pixel-wise classification problems is the cross-entropy loss:
\begin{equation}\label{eq:loss}
loss(\bm{f}) = -\frac{1}{N}\sum_{i=1}^{N}\log f_i(y_i^*) 
\end{equation}
where $N$ is the number of pixels in a batch, $y_i^*\in \{0, 1\}$ is the ground truth class of a pixel $i$. $y_i^* = 1$ means pixel $i$ is in the signal area. $f_i(y_i^*)$ is the predicted probability of pixel $i$ belongs to class $y_i^*$. The predicted class for a given pixel $i$ is $\hat{y_i} = \arg \max_{y \in \{0, 1\}} f_i(y)$.

\subsubsection{Lovasz hinge loss} 
Proposed by \cite{berman2018lovasz}, the Jaccard index, also referred to as the intersection-over-union score, is commonly used in the evaluation of semantic segmentation:
\begin{equation}\label{eq:IOU}
J_c(\hat{y}, y^*) = \frac{\left | \{ \hat{y} =  c\} \cap  \{ y^* =  c\}   \right |}{\left | \{ \hat{y} =  c\} \cup \{ y^* =  c\}   \right |}
\end{equation}
Then the miss detection is defined as:
\begin{equation}\label{eq:MD}
\Delta_{J_c} (\hat{y}, y^*) = 1 - J_c(\hat{y}, y^*)
\end{equation}
The ground truth $y^*$ is fixed. If we take the derivative of $\Delta_{J_c}$ with regard to the network output $F_i(c)$, the result is either 0 or inf, which makes it impossible to optimize. The $\Delta_{J_c}$ can be rewritten as a function of a set of mispredictions $m$, where the mispredictions is defined as:
\begin{equation}\label{eq:misp}
m_i = \left\{\begin{matrix}
0, &if & y_i^* = \hat{y_i}   \\ 
1, &if & y_i^* \neq  \hat{y_i}
\end{matrix}\right.
\end{equation}
Lovasz hinge loss is an interpolation version of $\Delta_{J_c}$ by:
\begin{equation}\label{eq:lhl}
\Bar{\Delta}_{J_c} (m_{\pi_i}) = \sum_{i=1}^{p} m_{\pi_i} g_{\pi_i}(\mathbf{m})
\end{equation}
where $p$ is the number of samples, $\pi$ is a permutation ordering of components of m in decreasing order, i.e. $m_{\pi_1} \geqslant m_{\pi_2} \ldots \geqslant m_{\pi_p}$. The ground truth label is mapped to $y^* \in \{ -1, 1\}$ and misprediction is replaced with the hinge loss for an input image \textbf{x}:
\begin{equation}\label{eq:hinge_loss}
m_{\pi_i} = \max (1 - F_{\pi_i}(\mathbf{x})y_{\pi_i}^* , 0) \in \mathbb{R}^p
\end{equation}
and the contribution of sample $\pi_i$ to $\Delta_{J_c}$ is given by 
\begin{equation} \label{eq:grad}
    g_{\pi_i}(\mathbf{m}) = \Delta_{J_c} (\{\pi_1,~\ldots,~\pi_i\}) - \Delta_{J_c} (\{\pi_1,~\ldots,~\pi_{i-1}\}).
\end{equation}
where $\{\pi_1,~\ldots,~\pi_i\}$ means only the first $\pi_i$ points with highest loss of $\hat{y}$ are used and $\hat{y}_j = y^*_j$ for $j > \pi_i$. Lovasz loss is differentiable with regard to the network output $F(\mathbf{x})$ and it is easy to be plugged into the segmentation model. 

\subsection{Training}
We initialize the weights and bias in our networks randomly and use Adam optimizer. The learning rate is $10^{-3}$ and the training epochs number is 50 with a batch size of 4. The images are normalized before sent to the network. The networks is determined by the kernels in each layer and is independent to the image size, which gives us the flexibility that we do not need to scale or crop the input image again when it comes to the testing time.

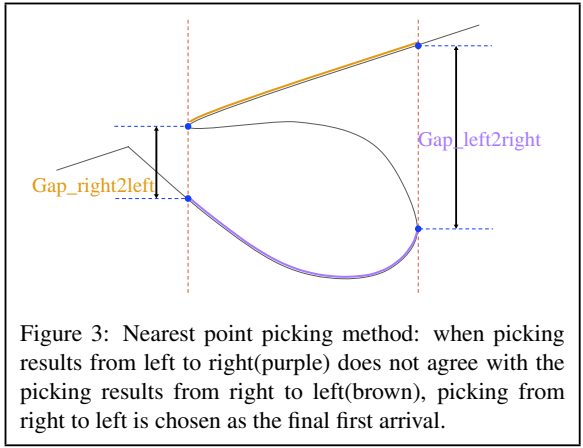
\plot{npp}{width=\columnwidth}
{Nearest point picking method: when picking results from left to right(purple) does not agree with the picking results from right to left(brown), picking from right to left is chosen as the final first arrival.
}

\subsection{Post-processing Picking}
\subsubsection{first point picking (FPP)}
Suppose we have $R$ receivers in one shot gather images. The simplest picking method is for each receiver $r$, pick the first pixel which is predicted to be a true signal.
\begin{equation}\label{eq:fpp}
\hat{t}_r = \arg \min_t \{t | \hat{y}_{(r, t)}=1\}, \quad r \in [0, R) 
\end{equation}
where $(r, t_r)$ are the coordinates of the selected pixel for receiver $r$.

\subsubsection{nearest point picking (NPP)}
For each receiver $r$, after we get the prediction map from the neural networks, we would get a prediction trace for this particular receiver. Since we know that the first arrival must on the edge, we can find all possible candidates by tracking the edge as follows:
\begin{equation}\label{eq:cand}
t_r \in \{ Candidates\}, \quad if \quad \hat{y}_{(r, t)} - \hat{y}_{(r, t-1)}=1
\end{equation}
After that, we use the correlation of the first arrivals across the traces to select the most reasonable candidate. Given the picking point from the adjacent trace $t_{r-1}$, the picking point in current trace is selected by 
\begin{equation}\label{eq:npp}
\hat{t}_r = \arg \min_{t_r} \left | t_r - t_{r-1} \right | \quad for \quad t_r \in \{ Candidates\}
\end{equation}

One-direction NPP could leads to a biased picking and deviate from the true path due to an imperfect segmentation. To solve the problem, we consider applying NPP from left to right and then from right to left. After getting two picking lines, we compare the difference. If they agree with each other, then the common first arrival is picked. When they disagree, we calculate the gap at the edge of common picking point. As showed in \autoref{fig:npp} The picking direction with a lower gap wins the competition as the error is likely to be accumulated during the nearest point picking. In other words, we conduct a greedy search on both directions and locate the local minimal based on the segmentation mask. 

\tabl{results}{Seismic image segmentation and picking results on the testing set. Rows 2-4 use cross-entropy loss while rows 5-7 use Lovasz loss (ts is time step)}
{
\begin{center}
    \resizebox{8cm}{!}{
        \begin{tabular}{|c|c|c|c|c|}
        \hline
        \textbf{Datasets} & \textbf{\begin{tabular}[c]{@{}c@{}}synthetic \\ data\end{tabular}} & \textbf{\begin{tabular}[c]{@{}c@{}}synthetic \\ Disc. data\end{tabular}} & \textbf{\begin{tabular}[c]{@{}c@{}}synthetic \\ noisy data\end{tabular}} & \textbf{\begin{tabular}[c]{@{}c@{}}field\\  data\end{tabular}} \\ \hline
        \textbf{\begin{tabular}[c]{@{}c@{}}Model \\ trained on\end{tabular}} & \begin{tabular}[c]{@{}c@{}}synthetic \\ data\end{tabular} & \begin{tabular}[c]{@{}c@{}}synthetic \\ data\end{tabular} & \begin{tabular}[c]{@{}c@{}}synthetic \\ data\end{tabular} & \begin{tabular}[c]{@{}c@{}}field\\  data\end{tabular} \\ \hline
        \textbf{\begin{tabular}[c]{@{}c@{}}Acc with \\ CE loss\end{tabular}} & 99.13\% & 99.09\% & 98.79\% & 97.83\% \\ \hline
        \textbf{\begin{tabular}[c]{@{}c@{}}MAE with \\ FPP (ts)\end{tabular}} & 22.56 & 23.50 & 26.84 & 27.54 \\ \hline
        \textbf{\begin{tabular}[c]{@{}c@{}}MAE with \\ NPP (ts)\end{tabular}} & 0.85 & 4.27 & 11.66 & 11.52 \\ \hline
        \textbf{\begin{tabular}[c]{@{}c@{}}Acc with \\ Lovasz loss\end{tabular}} & 99.92\% & 99.91\% & 99.18\% & 98.97\% \\ \hline
        \textbf{\begin{tabular}[c]{@{}c@{}}MAE with \\ FPP (ts)\end{tabular}} & 0.68 & \textbf{0.79} & \textbf{1.65} & 5.55 \\ \hline
        \textbf{\begin{tabular}[c]{@{}c@{}}MAE with \\ NPP (ts)\end{tabular}} & \textbf{0.60} & \textbf{0.79} & 2.33 & \textbf{4.89} \\ \hline
        \textbf{Image size} & \begin{tabular}[c]{@{}c@{}}1250 * \\ 2000\end{tabular} & \begin{tabular}[c]{@{}c@{}}1250 *\\  (1400$\sim$2000)\end{tabular} & \begin{tabular}[c]{@{}c@{}}1250 * \\ 2000\end{tabular} & \begin{tabular}[c]{@{}c@{}}750 * \\ (600$\sim$700)\end{tabular} \\ \hline
        \end{tabular}
    }
\end{center}
}

\section{Experiments}
In this section, we evaluate our method on three datasets: synthetic data, noisy and disconnected synthetic data, and field data. Our first arrival picking algorithm performs well on all of them. The results are summarized in \autoref{tbl:results}. All of the models have the same structure. We can find that the best model is Unet with Lovasz loss and NPP method.

\subsection{Metrics}
Semantic segmentation can also be considered as a pixel-wise classification task. Thus we can use classification accuracy to measure the performance of our convolutional network. The segmentation accuracy is
\begin{equation}\label{eq:accu}
accuracy(\bm{f}) = \frac{1}{mn}\sum_{j=1}^{m}\sum_{i=1}^{n} I(\hat{y}_{ij} = y_{ij}^* | f_{ij}) 
\end{equation}
where $m$ is the number of images in a validation set or test set, $n$ is the number of pixels in one image, $I(X)$ is the indicator function, which means it equals 1 if $X$ is true, otherwise 0. $f_{ij}$ is the network prediction for image $j$ at pixel $i$.

The aforementioned metric alone is not enough since the main focus is the boundary and the classification results of the other pixels are not that important. To measure the boundary directly, we use mean absolute error(MAE)
\begin{equation}\label{eq:MAE}
MAE(\bm{\hat{t}}) = \frac{1}{mR}\sum_{j=1}^{m}\sum_{r=1}^{R} \left | \hat{t}_{jr} - t_{jr}^* \right |
\end{equation}
where $R$ is the number of columns in one image, $\hat{t}_{jr}$ is the predicted arrival time of the first signal received by receiver $r$ at shot gather $j$, $t_{jr}^*$ is the ground truth for that signal. MAE is the core metric we use to measure the performance of our first arrival picking algorithm, the lower the better.

\plot{syn_noisy_ori}{width=\columnwidth}
{Compare the results of Lovass loss with NPP and CE loss with FPP on a synthetic noisy data. Red line is the ground truth, blue is the picking by the model, the harmonic noise is denoted in the orange box.
}

\plot{field_change_mask}{width=\columnwidth}
{(a) is the ground truth mask for this concatenated shot gathers from field; (b) is the segmentation mask given by the model with Lovasz loss; (c)is the picking results of FPP; (d) is the picking results of NPP. 
}

\subsection{Synthetic seismic data}
The first experiment is tested on the synthetic seismic data. The image size is 1250*2000. The sample rate is 8ms. We use cropped images with the width of 600 to train the model. We use 1000 samples for training and 200 samples for validation. The lowest MAE is 0.6, and the average timestamp error is only 4.8ms.

\subsection{Noisy and disconnected synthetic data}
Instead of adding simple incoherent noises such as Gaussian white noise or salt and pepper noise, we performed the spectrum analysis and designed a noise generator to synthesize the multi-harmonic noises that are very typical observed in many field seismic datasets.  The amplitude of the noise is modeled by Gaussian random process to simulate the continuity across the traces. The maximal amplitude of noise is set to be half of the maximum signal amplitude. This noise generator was designed based on our observation and analyses of large volume of the field datasets; hence we believe it produces very realistic multi-harmonic noises. In addition, we may also have some images with disconnected first arrival signals due to the missing receivers or spliced shot gathers. \autoref{fig:syn_noisy_ori} proves our model performs well with noisy data or traces with gaps.

\subsection{Field seismic data}
Field data is more complicated compared to synthetic data. The raw data has different size and each image contains several shot gathers. Almost every shot gather is contaminated by the noises. Each field image has a size around 750 * 650. \autoref{fig:field_change_mask} shows that when the segmentation results has mispredictions in the upper area, some noises will be picked as the first arrival if we use FPP method. In contrast, if we use NPP method, the mispredicted picking points will be "pulled down" and the picking results are much close to the groundtruth despite segmentation errors. The algorithm also has the ability to pick small shot gather with width less than 5. The model can be further improved with more training data.

\section{Conclusion}
In this paper, we proposed an improved deep learning segmentation method plus a parameter-free post-processing picking method to automatically pick the first arrival signals for the seismic data. We uses the latest surrogate IoU loss, Lovasz loss, to train our model. Compared to the cross entropy loss, we proved that Lovasz loss can improve the overall performance. We also test our method with noisy data. Finally, the reliability of the model is validated on field data with high accuracy compared to hand-picking results.

\section{Acknowledgments}

This material is based upon work supported by National Science Foundation under Grant No. 1746824

\onecolumn

\bibliographystyle{seg}  
\bibliography{bib/example}

\end{document}